\documentclass[12pt,a4paper]{elsarticle}
\usepackage{amsmath}
\usepackage{booktabs}
\usepackage{graphicx}
\usepackage[pdfauthor={Wanja Timm Schulze}]{hyperref}
\usepackage{listings}
\usepackage{microtype}
\usepackage{subcaption}
\usepackage{xcolor}

\journal{\href{https://doi.org/10.1016/j.softx.2025.102035}{SoftwareX}}
\date{January 15, 2025}
\newcommand{\eminus}{\textsc{\mbox{eminus}}}
\begin{document}

\begin{frontmatter}

\title{eminus --- Pythonic electronic structure theory}
\author[label1]{Wanja Timm Schulze\corref{cor1}}
\ead{wanja.schulze@uni-jena.de}
\address[label1]{Institute for Physical Chemistry, Friedrich Schiller University, 07743 Jena, Germany}
\cortext[cor1]{Corresponding author}
\author[label2,label3]{Sebastian Schwalbe}
\address[label2]{Center for Advanced Systems Understanding, 02826 Görlitz, Germany}
\address[label3]{Helmholtz-Zentrum Dresden-Rossendorf, 01328 Dresden, Germany}
\author[label4]{Kai Trepte}
\address[label4]{Taiwan Semiconductor Manufacturing Company North America, San Jose, USA}
\author[label1]{Stefanie Gräfe}


\begin{abstract}
In current electronic structure research endeavors such as warm dense matter or machine learning applications, efficient development necessitates non-monolithic software, providing an extendable and flexible interface.
The open-source idea offers the advantage of having a source code base that can be reviewed and modified by the community.
However, practical implementations can often diverge significantly from their theoretical counterpart.
Leveraging the efforts of recent theoretical formulations and the features of Python, we try to mitigate these problems.
We present \eminus{}, an education- and development-friendly electronic structure package designed for convenient and customizable workflows, yet built with intelligible and modular implementations.
\end{abstract}

\end{frontmatter}

\begin{table*}[htbp]
    \caption{Code metadata.}
    \footnotesize
    \begin{tabular}{|l|p{6.5cm}|p{5.3cm}|}
        \hline
        \textbf{Nr.} & \textbf{Code metadata description} & \textbf{Code metadata} \\
        \hline
        C1 & Current code version & 3.0.1 \\
        \hline
        C2 & Permanent link to code/repository used for this code version & \url{https://github.com/wangenau/eminus} \\
        \hline
        C3 & Permanent link to Reproducible Capsule & \url{https://hub.docker.com/r/wangenau/eminus} \\
        \hline
        C4 & Legal Code License & Apache-2.0 \\
        \hline
        C5 & Code versioning system used & Git \\
        \hline
        C6 & Software code languages, tools, and services used & Python \\
        \hline
        C7 & Compilation requirements, operating environments \& dependencies & Python\,\textgreater=\,3.7, NumPy\,\textgreater=\,1.17, SciPy\,\textgreater=\,1.6 \\
        \hline
        C8 & Link to developer documentation/manual & \url{https://wangenau.gitlab.io/eminus} \\
        \hline
        C9 & Support email for questions & \href{mailto:wangenau@protonmail.com}{wangenau@protonmail.com} \\
        \hline
    \end{tabular}
\end{table*}

\section{Motivation and significance}
The properties of materials, such as band structures, ionization potentials, or dipole moments, are determined by their electronic structure, which are theoretically described with the help of various quantum mechanical methods.
This can be achieved by finding solutions to the Schrödinger equation of the system of interest, which however can only be solved analytically for simple systems; solving it is generally a non-trivial problem.
As a result, approximations are necessary to perform material investigations.
A set of these approximations \cite{Hohenberg1964, Kohn1965} led to the well-known and widely used density functional theory (DFT).

Due to its moderate computational cost compared to other quantum theory methods \cite{Friesner2005}, such as wave function methods, DFT has become one of the most commonly employed electronic structure methods \cite{vanMourik2014} for studying atoms, molecules, clusters, condensed matter \cite{Kohn1999}, and warm dense matter (WDM)~\cite{BenuzziMounaix2014}.
While DFT is formally exact \cite{Hohenberg1964}, the exchange-correlation functional that describes the many-body electron-electron interactions in DFT remains unknown.
This has led to the development of density functional approximations (DFAs) to estimate the exchange-correlation interactions.
DFAs are separated into different levels of theory with increasing complexity~\cite{Perdew2001}, allowing researchers to tailor calculations to specific needs and computational resources.

To represent the wave function as the solution of the Schrödinger equation, a finite basis set is typically used.
For solid-state and WDM calculations, plane waves \cite{Payne1992} are a common choice of the basis.
Calculations employing plane waves, among other basis sets, can be carried out using different electronic structure packages, such as the commonly utilized open-source programs Abinit \cite{Gonze2020}, CP2K \cite{Kuehne2020}, DFTK \cite{Herbst2021}, GPAW \cite{Mortensen2005}, JDFTx \cite{Sundararaman2017}, PWDFT.jl~\cite{Fathurrahman2020}, KSSOLV \cite{Yang2009}, or QUANTUM ESPRESSO (QE) \cite{Giannozzi2009}.
Another popular basis set choice for solid-state calculations employs finite-difference grids, implemented in programs such as DFT-FE \cite{Motamarri2020}, M-SPARC \cite{Xu2020}, SPARC \cite{Xu2021}, or Octopus \cite{TancogneDejean2020}.
More available options can be found in Ref. \cite{Lehtola2022}.

Recent advancements, such as finite temperature exchange-correlation functionals \cite{Karasiev2014, Groth2017, Karasiev2018, Dornheim2018, Mihaylov2020, Karasiev2022} and specialized solvers \cite{Suryanarayana2018, White2020} in WDM studies, or the rapid progress of machine learning (ML), such as ML functionals \cite{Kirkpatrick2021, Pederson2022, Voss2024, Kelley2024}, are in need of non-monolithic DFT codes with flexible interfaces.
Open-source programs are particularly advantageous in this context by having a code base that can be reviewed, modified, and extended, enabling the research community to improve existing theories and implement new ones directly.

However, in the end, theoretical formulations and practical implementations of a theory can often diverge significantly.
Consequently, understanding a method to its full extent can become extensively complex.
To address this issue, efforts have been made to express DFT in an algebraic formulation, titled DFT\raisebox{0.125em}{\footnotesize++} \cite{IsmailBeigi2000}, that closely resembles its software counterpart.

The programming language also plays a crucial role in the accessibility of electronic structure software.
Python \cite{vanRossum2009} is one of the most popular programming languages \cite{TIOBE2024} that also showed success in other electronic structure packages like PySCF \cite{Sun2020}, GPAW \cite{Mortensen2005}, HORTON \cite{Chan2024}, or PyQuante2~\cite{Muller2017}.
Its high-level abstraction allows writing accessible and readable code closely resembling the theoretical formulation.
Furthermore, by leveraging powerful linear-algebra libraries like NumPy \cite{Harris2020}, Python allows for writing concise and performant code.
Building on these strengths, we introduce the \eminus{} software package which utilizes the algebraic DFT\raisebox{0.125em}{\footnotesize++} formulation to implement DFT with a plane wave basis set from scratch.

The structure of this article is as follows: Section \ref{sec:software} describes the software design of \eminus{}, including the translation of theoretical formulations, development details, and software functionalities.
Selected illustrative examples are presented in section \ref{sec:examples}.
The article concludes with the impact of the described software and concluding remarks in section \ref{sec:conclusion}.

\section{Software description}
\label{sec:software}
\subsection{Software architecture}
\eminus{} is built as a Python package, combining functional and object-o\-ri\-ented programming.
The heart of the project are the \lstinline|Atoms| and \lstinline|SCF| classes.
The \lstinline|Atoms| class contains atomic information as well as the simulation cell and basis set setup.
To perform self-consistent field (SCF) calculations, such as DFT calculations, the \lstinline|SCF| class is utilized.
It contains the calculation parameters, like the exchange-correlation functional, but also stores results, such as the energy contributions in an \lstinline|Energies| data class.

By employing the algebraic formulation of DFT\raisebox{0.125em}{\footnotesize++}, the core \lstinline|dft| module ensures a readable and straightforward implementation of DFT.
A prominent example to showcase this is solving the Poisson equation.
In the operator notation of DFT\raisebox{0.125em}{\footnotesize++}, the equation reads \cite{IsmailBeigi2000}
\begin{align}
    \phi(\boldsymbol r) = -4\pi\mathcal L^{-1}\mathcal O\mathcal J n(\boldsymbol r),
    \label{eq:poisson}
\end{align}
with $\phi$ being the Hartree field induced by the mean electron density $n$ depending on the coordinate $\boldsymbol r$, and $\mathcal L^{-1}$, $\mathcal O$, and $\mathcal J$ being basis set-dependent operators.
These include the inverse Laplacian operator $\mathcal L^{-1}$, the overlap operator $\mathcal O$, and the inverse transformation operator $\mathcal J$.
Notably, the formulation of Eq. \eqref{eq:poisson} is independent of the basis set, though the formulation (and implementation) of the basis set-dependent operators will differ depending on the basis set.

With the \lstinline|Atoms| class holding the basis set-dependent properties, the actual implementation can be seen in Lst. \ref{lst:poisson} (excluding documentation).
\begin{figure*}[htbp]
\begin{lstlisting}[caption={Implementation of calculating a solution to the Poisson equation.},label={lst:poisson}]
def get_phi(atoms, n):
    return -4 * np.pi * atoms.Linv(atoms.O(atoms.J(n)))
\end{lstlisting}
\end{figure*}

When it comes to more complex expressions, one can use more of Python's language features to write more comprehensible code.
One example is the orthogonalization of the wave functions $W$, stored in the coefficient matrix \lstinline|W|.
In the formulation of DFT\raisebox{0.125em}{\footnotesize++}, the orthogonalized wave functions $Y$ are calculated via \cite{IsmailBeigi2000}
\begin{align}
    Y = W \left(W^\dag \mathcal OW\right)^{-1/2}.
    \label{eq:ortho}
\end{align}

The practical implementation needs to be generalized over multiple $\boldsymbol k$-points \cite{Payne1992} and spin channels, resulting in the function displayed in Lst. \ref{lst:orth_naive} (again excluding any documentation).
\begin{figure*}[htbp]
\begin{lstlisting}[caption={Naive implementation of a wave function orthogonalization procedure.},label={lst:orth_naive}]
def orth(atoms, W):
    Y = [np.empty_like(Wk) for Wk in W]
    for ik in range(atoms.kpts.Nk):
        for spin in range(atoms.occ.Nspin):
            Y[ik][spin] = (W[ik][spin]
                           @ inv(sqrtm(W[ik][spin].conj().T
                           @ atoms.O(W[ik][spin]))))
    return Y
\end{lstlisting}
\end{figure*}

Compared to Eq. \eqref{eq:ortho}, this implementation can be understood on its own given the theoretical formulation, especially through Python's matrix multiplication operator \lstinline|@|.
However, the use of many nested blocks to handle different \(\boldsymbol k\)-points and spin channels increases the cyclomatic complexity \cite{McCabe1976, Ebert2016}, a quantitative measure of the complexity of a program.

To reduce the cyclomatic complexity, \eminus{} uses Python decorators.
Decorators can be used to wrap functions, allowing one to modify their behavior efficiently.
In the aforementioned case, the function can be simplified with two reusable custom decorators, resulting in an implementation with reduced cyclomatic complexity (from 4 to 1, compared to Lst. \ref{lst:orth_naive}, calculated with Radon \cite{Lacchia2012}) while closely resembling the theoretical equation, as seen in Lst.~\ref{lst:orth}.
\begin{figure*}[htbp]
\begin{lstlisting}[caption={Implementation of a wave function orthogonalization procedure using decorators.},label={lst:orth}]
@handle_k
@handle_spin
def orth(atoms, W):
    return W @ inv(sqrtm(W.conj().T @ atoms.O(W)))
\end{lstlisting}
\end{figure*}

\subsubsection{Distribution}
The source code of \eminus{} is hosted on GitHub and GitLab, while the Python package is distributed via PyPI.
It can be installed like any Python package, using tools such as pip \cite{PPA2009}, pipx \cite{PPA2019}, or uv \cite{Astral2024}.
The core functionalities of the package solely depend on NumPy and SciPy \cite{Virtanen2020}, simplifying the management of dependencies.
For the fully reproducible usage under NixOS~\cite{Kowalewski2022}, a developer shell is provided as well.

To streamline the development process, continuous in\-tegration/continuous delivery (CI/CD) pipelines automate various tasks, including building the package.
Each release generates a Docker container, with the build and attestation process for both the Python package and Docker container automated and publicly visible using the CI/CD capabilities of GitLab \cite{GitLab2024}.

\subsubsection{Testing}
To ensure consistent behavior, an extensive test suite has been built that tests either for reference energies, physical properties, or the expected behavior of classes and functions.
The test suite can be invoked using pytest \cite{Krekel2004} and is automatically executed via GitLab's CI/CD.
Additional code coverage reports are generated for every release to monitor the portion of executed code in the tests, with the current coverage rate being approximately 97\%.

Since Python is dynamically typed, \eminus{} uses type hints using so-called stub files for improved error detection, documentation, and code completion in integrated development environments (IDEs).
Strict type check tests are carried out with mypy \cite{Lehtosalo2012}, again, via GitLab's CI/CD.

For broad platform support, tests are conducted across all supported Python versions and various platforms, including Ubuntu, Debian, macOS, Windows, and NixOS.
To catch stylistic errors and to maintain a readable and consistent source code, the project uses Ruff \cite{Astral2022} for linting and formatting.

\subsubsection{Documentation}
To make the usage of \eminus{} easy-to-pick-up, a web documentation is generated using Sphinx \cite{SDT2007}.
This documentation includes installation instructions, extensive examples, automated module descriptions, and developer information.
The developer information covers changelogs, license details, descriptions of the most commonly used variables in the code, and how the development tools can be utilized.

If problems are encountered, users can open issues online in the repository.
In addition, a Discord server \cite{Schulze2024a} has been set up to provide straightforward support.

\subsection{Software functionalities}
In its core, \eminus{} is a plane wave DFT code, supporting $\boldsymbol k$-point-dependent calculations.
Both restricted and unrestricted formulations are available, allowing efficient spin treatment, such as setting spin states or initial magnetizations.
SCF and band structure calculations supporting fixed occupations and Fermi-smearing \cite{Gillan1989} are implemented.
DFT calculations can be performed with local density approximation (LDA), generalized gradient approximation (GGA), and \textit{meta}-GGA exchange-correlation functionals \cite{Perdew2001}.
While a handful of functionals are directly implemented, an interface to Libxc \cite{Lehtola2018} is provided as well.
Further, a flexible interface is provided to modify the parameters of internal and Libxc functionals, such as adjusting the temperature in thermal exchange-correlation functionals.

By default, the norm-conserving GTH pseudopotentials \cite{Goedecker1996} are used in calculations to replace the electron-ion interaction.
The minimization of the DFT energy is done via direct minimization \cite{Teter1989, IsmailBeigi2000} with the option to customize optimization schemes, such as by mixing steepest descent, line minimization, and conjugate gradient methods.

\eminus{} supports XYZ, POSCAR, and CUBE files to input atom symbols and positions.
Additionally, PDB structures can be generated.
All class objects can be stored in JSON or HDF5 files to save and load them later or to restart any calculation.

Many properties and measures have been implemented in the \eminus{} toolbox, including electron localization functions \cite{Becke1990}, reduced density gradients (RDGs) \cite{Perdew1996}, ionization potentials, magnetizations, and the generation of different types of orbitals, to name a few.
The latter comprises of Kohn--Sham, Wannier \cite{Silvestrelli1999}, selected columns of the density matrix (SCDM) \cite{Damle2015}, Fermi, and Fermi--Löwdin orbitals (FLOs) \cite{Pederson2014}.
Except for the Kohn--Sham orbitals, the generations are restricted to the $\Gamma$-point for now.
Self-interaction correction (SIC) \cite{Perdew1981} functionalities can be utilized using said orbitals.
It should be noted that force and stress calculations have not yet been implemented.

Additional functionalities can be accessed through modular extras that require optional dependencies like PySCF \cite{Sun2020}.
These extras include DFT-D3 dispersion corrections \cite{Grimme2010}, Fermi-orbital descriptor (FOD) generation \cite{Schwalbe2019}, HDF5 file reading and writing \cite{Collette2013}, an interface to Libxc \cite{Lehtola2018}, $\boldsymbol k$-point symmetrization, faster operators using Torch \cite{Paszke2019}, and visualization features.
The visualization extra, utilizing NGLView \cite{Nguyen2018} and Plotly \cite{PlotlyTechnologies2015}, allows users to display the simulation cell, isosurfaces, contour lines, various file types, band structures, and Brillouin zones with $\boldsymbol k$-points.
The visualization features will be exemplified in Sec. \ref{sec:examples}.
To ensure compatibility with open-source principles, all dependencies of \eminus{} have licenses approved by the Open Source Initiative \cite{OSI2024}.

\subsection{Benchmarks}
\label{subsec:benchmarks}
To demonstrate the comparability of \eminus{} with other electronic structure packages, a series of DFT calculations has been conducted.
The codes to compare against are JDFTx (written in C\raisebox{0.125em}{\footnotesize++}), PWDFT.jl (Julia), and QE (Fortran).

\begin{figure*}[htbp]
    \centering
    \includegraphics[width=\textwidth]{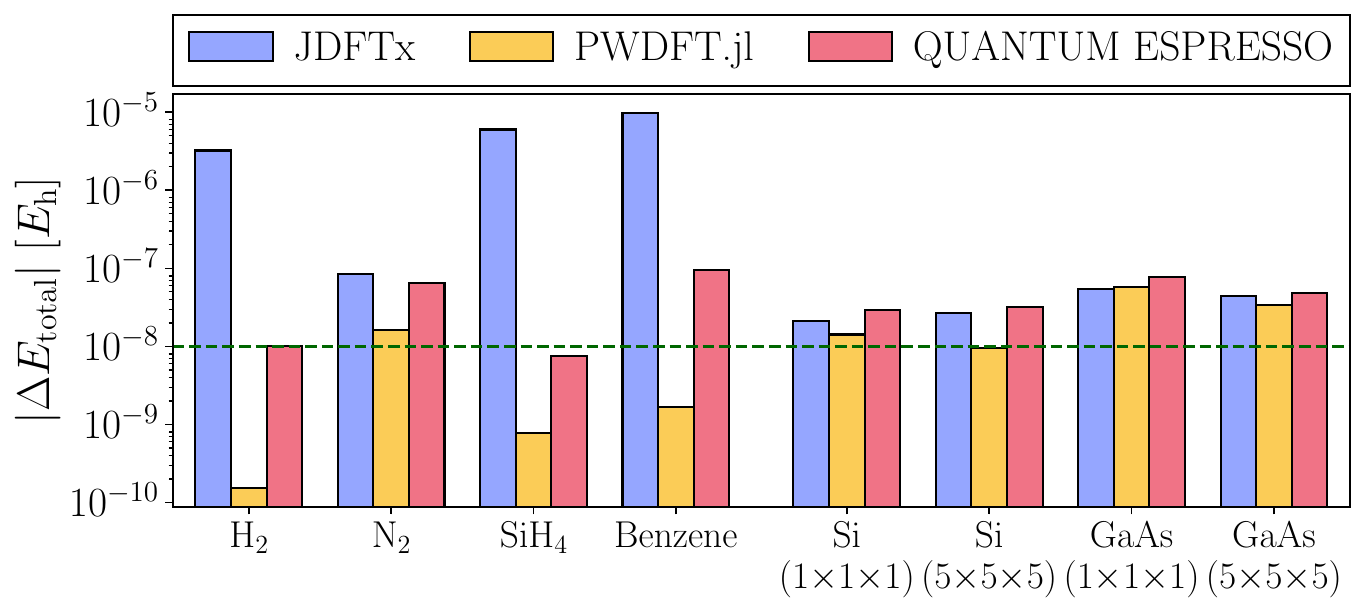}
    \caption{Absolute total energy differences $|\Delta E_\mathrm{total}| = |E^\mathrm{\eminus{}}_\mathrm{total} - E^\mathrm{code}_\mathrm{total}|$ in $E_\mathrm{h}$ for DFT calculations using different codes, compared to \eminus{}, visualized with Matplotlib \cite{Hunter2007}.
    All calculations use the SVWN5 exchange-correlation functional, a cut-off energy of $30\,E_\mathrm{h}$, and a total energy convergence threshold of $10^{-8}\,E_\mathrm{h}$.
    For solids, the $\boldsymbol k$-point grid is given in parentheses.
    The energy convergence threshold is marked as a dashed horizontal line.}
    \label{fig:energies}
\end{figure*}

The total energies of these DFT calculations are analyzed for a set of mol\-e\-cules and solids.
All calculations employed the Slater exchange \cite{Bloch1929, Dirac1930} and the VWN correlation functional \cite{Vosko1980} (SVWN5), a cut-off energy of $30\,E_\mathrm{h}$, and a total energy convergence threshold of $10^{-8}\,E_\mathrm{h}$.
The molecule calculations use a $20\,a_0$ unit cell and are $\Gamma$-point-only while the solids utilized different Monkhorst--Pack $\boldsymbol k$-point grids \cite{Monkhorst1976} (see Fig. \ref{fig:energies}).
Resulting absolute total energy differences compared to \eminus{} are shown in Fig. \ref{fig:energies} (see Tab. \ref{tab:energies} for a listing of these values).

From this comparison, it is evident that even for vastly different implementations using diverse programming languages, all codes produce almost identical energies.
This is especially noteworthy since all codes use different optimization schemes with distinct numerical parameters.
Even for a very tight threshold, the energy difference is often below the convergence threshold but always close to it.
The best agreement can be found with PWDFT.jl.

Additional benchmark details, including input files, calculation logs, software versions, and used hardware are provided in the supplementary repository to this article \cite{Schulze2024}.
Performance comparisons for these calculations are discussed in the appendix in Sec. \ref{sec:timings}.

\section{Illustrative examples}
\label{sec:examples}
Unlike most electronic structure software \cite{Oboyle2007}, \eminus{} does not rely on input files or output logs.
Instead, it can be used as a Python package in calculation scripts.
This enables the creation of complex workflows, direct access to calculation parameters and results, and encourages interactive exploration, such as in Jupyter notebooks \cite{Kluyver2016}.
To illustrate the user-friendliness of \eminus{}, we will discuss two example calculations.

\subsection{Example 1: Bulk silicon}
As a standard textbook example, Lst. \ref{lst:silicon} showcases a calculation for bulk silicon.
First, a cell for silicon in diamond configuration with a lattice parameter of $10.263\,a_0$ is generated.
Here, \lstinline|Cell| is a convenience wrapper to create a suited \lstinline|Atoms| object for crystal structures.
Additionally, a cut-off energy of $30\,E_\mathrm{h}$, 8 electronic bands, a 3$\times$3$\times$3 Monkhorst--Pack $\boldsymbol k$-point grid, and a Fermi-smearing temperature of $0.005\,E_\mathrm{h}$ are set.

Afterward, an \lstinline|SCF| object is created using the PBE GGA exchange-correlation functional \cite{Perdew1996}.
After completing the SCF calculation, a new $\boldsymbol k$-point grid is generated by giving a band path with 50 sampling points.
The band energies are then minimized for the fixed Hamiltonian obtained from the prior DFT calculation.
Finally, the Brillouin zone, including special points and band path, as well as the band structure, can be displayed using the visualization extra.
The resulting figures are shown in Fig. \ref{fig:silicon}.

\begin{figure*}[htbp]
\begin{lstlisting}[caption={Input script for a bulk silicon band structure calculation.},label={lst:silicon}]
from eminus import Cell, SCF
from eminus.extras import plot_bandstructure

cell = Cell(atom="Si", lattice="diamond", ecut=30, a=10.263,
            bands=8, kmesh=(3, 3, 3), smearing=5e-3)
scf = SCF(cell, xc="pbe")
scf.run()

scf.kpts.path = "LGXU,KG"
scf.kpts.Nk = 50
scf.converge_bands()
scf.kpts.view()
plot_bandstructure(scf)
\end{lstlisting}
\end{figure*}

\begin{figure*}[htbp]
    \centering
    \begin{subfigure}{\textwidth}
        \raisebox{1.5em}{\includegraphics[width=0.42\textwidth]{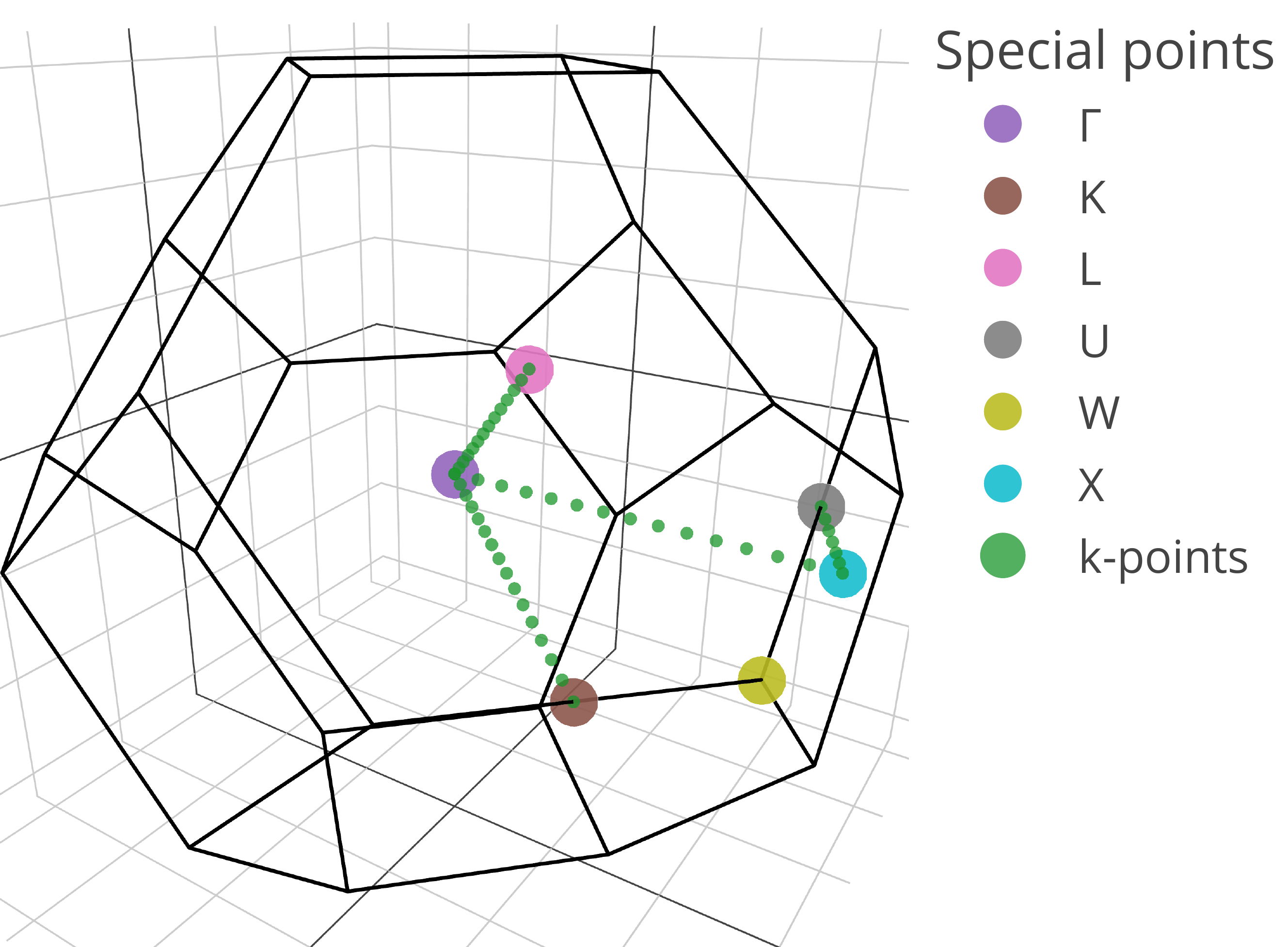}}
        \includegraphics[width=0.57\textwidth]{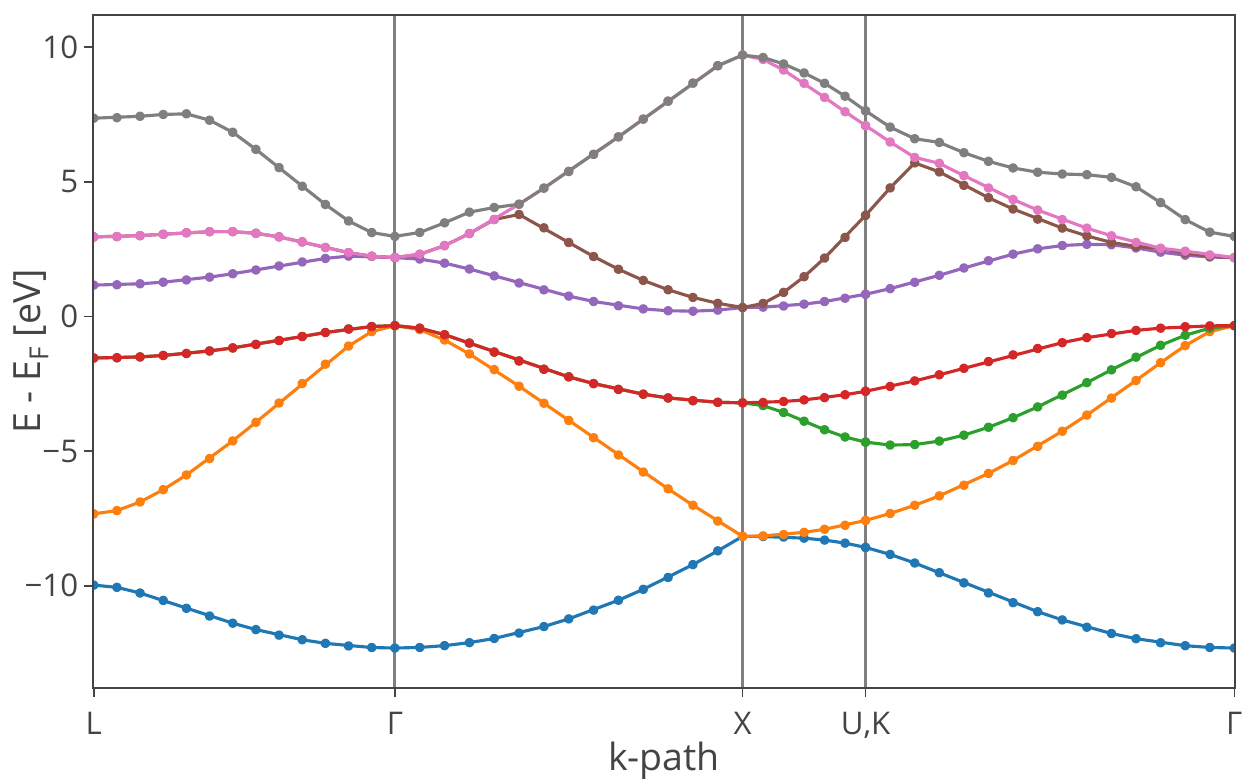}
    \end{subfigure}
    \caption{Brillouin zone (left) and band structure (right) for bulk silicon, as generated by the Python script in Lst. \ref{lst:silicon}.
    Note that the colors of the special points in the Brillouin zone have no association with the colors in the band structure.}
    \label{fig:silicon}
\end{figure*}

\subsection{Example 2: Warm dense matter}
To illustrate a more advanced workflow, we adapt a calculation from a recent publication.
The example involves generating a plot of the RDG $s[n]$ over the normalized electron density $n/n_0$ (see Fig. 2 in Ref. \cite{Moldabekov2024}), used to indicate the breaking of bound states in warm dense hydrogen (for more information, see Ref. \cite{Moldabekov2024}).

For that, the workflow in Lst. \ref{lst:wdm} involves a DFT calculation for a set of hydrogen atoms, with coordinates and simulation cell given in a POSCAR file.
Computational parameters include a cut-off energy of $810\,$eV, 28 electronic bands, a 2$\times$2$\times$2 Monkhorst--Pack $\boldsymbol k$-point grid, and a Fermi-smearing temperature of $12.528\,$eV.
The SCF calculation converges the total energy until the difference is below $10^{-4}\,$eV, using the thermal GDSMFB LDA exchange-correlation functional \cite{Groth2017} as implemented in \eminus{}, where the temperature can be set as an external parameter, combined with a bare Coulomb potential.\footnote{The computational parameters have been reduced from the original publication, which uses a cut-off energy of $2700\,$eV, 280 bands, and a 10$\times$10$\times$10 $\boldsymbol k$-point grid.}

For hard-to-converge systems, it can be beneficial to tune the optimization procedure.
In this example, 25 steepest-descent steps will be done before performing the preconditioned conjugate gradient steps afterward.
Then, the electron density will be normalized and the RDG calculated.
The final RDG is visualized for an isovalue of $0.5\,a_0^{-3}$ and plotted against the normalized electron density as shown in Fig. \ref{fig:wdm}.

\begin{figure*}[htbp]
\begin{lstlisting}[caption={Input script for a DFT calculation of warm dense hydrogen.},label={lst:wdm}]
import matplotlib.pyplot as plt
import numpy as np
from eminus import Atoms, SCF, read, tools, units

sym, pos, cell = read("POSCAR")
atoms = Atoms(sym, pos, ecut=units.ev2ha(810), a=cell)
atoms.kpts.kmesh = 2
atoms.occ.bands = 28
atoms.occ.smearing = units.ev2ha(12.528)
scf = SCF(atoms, etol=units.ev2ha(1e-4), xc="LDA_XC_GDSMFB",
          pot="coulomb", opt={"sd": 25, "pccg": 250})
scf.xc_params = {"T": units.ev2ha(12.528)}
scf.run()

n = scf.n / (3 / (4 * np.pi * 2**3))
s = tools.get_reduced_gradient(scf)
scf.view(plot_n=s, isovalue=0.5)
plt.axhline(y=0.5)
plt.scatter(n, s, c=s)
\end{lstlisting}
\end{figure*}

\begin{figure*}[htbp]
    \centering
    \includegraphics[scale=0.76]{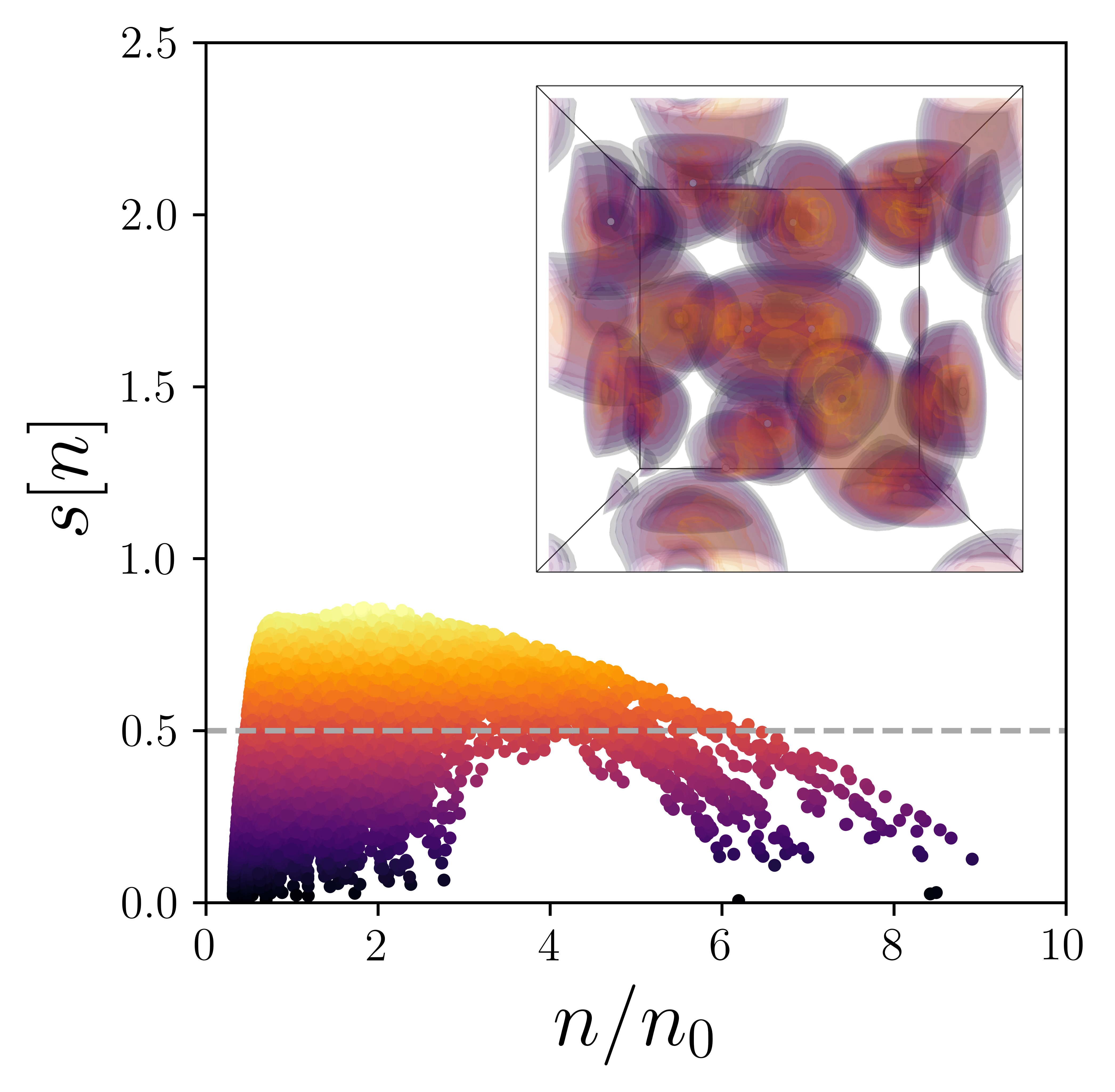}
    \caption{Reduced density grid $s[n]$ over the normalized electron density $n/n_0$ for a system of warm dense hydrogen, as generated by the Python script in Lst. \ref{lst:wdm}.
    The subplot shows the distributions of $s[n]$ in the real space (compare with Fig. 2 in Ref. \cite{Moldabekov2024}).}
    \label{fig:wdm}
\end{figure*}

It should be noted that the plot instructions were slightly modified to create the figures in this article.
Reproducible variants of the examples above can be found in the supplementary repository to this article \cite{Schulze2024}.
Additional examples, including simple geometry and FOD optimizations, Wannier orbital localization, functional parameterization, and more, can be found in the documentation of \eminus{}.

\section{Impact and conclusions}
\label{sec:conclusion}
When implementing new theories in computational chemistry or physics, the entry hurdle becomes quite large.
From learning the theory itself, new input scripts, low-level programming languages, and build tools can impede trying out new ideas and approaches.
Additionally, recent research and method development is in need of modifiable and extendable software.
Having a developer-friendly application, either for research or teaching purposes, can prove helpful in both cases.

In this article, we introduced \eminus{}, an electronic structure toolbox written in Python designed to be a user- and developer-friendly software package.
Combining the strengths of the algebraic formulation of density functional theory and the language features of Python allowed us to write source code that is very close to its theoretical formulation.
This easy-to-read source code provides reproducible implementations of electronic structure methods that can be used in custom-tailored Python workflows.
Additionally, \eminus{} offers broad platform support and rich documentation, further lowering the barriers to developing theoretical methods.
Given the previously mentioned aspects, we expect \eminus{} to be easy to pick up for future developments in teaching and research.

The utility of \eminus{} has already been exemplified in the working group of the authors where \eminus{} has been used for visualization and post-proc\-ess\-ing purposes \cite{Schulze2023, Schwalbe2024}.
Others used \eminus{} for the creation of reference data~\cite{Moldabekov2024}.
Overall, this led to frequent usage with over 600 downloads per month.\footnote{Averaged data from PyPI Stats \cite{Flynn2018} over the last 180 days, excluding mirrors.}

Future developments of \eminus{} will focus on improving its overall performance.
Support for a tensor computation framework like Torch or JAX \cite{Bradbury2018} can be introduced, e.g., via the array API standard \cite{ConsortiumMembers2023}.
Using Torch tensors can exhibit a speedup of factor two or more over NumPy arrays, as shown in a toy code of \eminus{} called SimpleDFT \cite{Schulze2021}.
Another way to enhance performance is by using GPU hardware support.
Besides performance gains, those frameworks would facilitate the implementation of methods such as automatic differentiation (AD).
Using AD would further accelerate systematic developments, as only the energy expressions specific to a given electronic structure method need to be implemented, allowing gradients to be calculated automatically \cite{Tan2023}.

\section*{Acknowledgments}
W.\,T.\,S. thanks Dr.\,Alexander Croy for helpful discussions regarding this article and Dr.\,Jens Kortus for support in the early conceptualization stage of the project during a master thesis.
W.\,T.\,S. and S.\,G. highly acknowledge funding by the Deutsche Forschungsgemeinschaft (DFG, German Research Foundation) -- Research unit FuncHeal, project ID 455748945 -- FOR 5301 (project P5).
S.\,S. was supported by the European Research Council (ERC) under the European Union's Horizon 2022 Research and Innovation Program (Grant Agreement No. 101076233, ``PREXTREME'') and by the Center for Advanced Systems Understanding (CASUS), financed by Germany's Federal Ministry of Education and Research (BMBF) and the Saxon State Government out of the State Budget approved by the Saxon State Parliament.
Views and opinions expressed are however those of the authors only and do not necessarily reflect those of the European Union or the European Research Council Executive Agency.
Neither the European Union nor the granting authority can be held responsible for them.

\biboptions{sort&compress}
\bibliographystyle{elsarticle-num}
\bibliography{references.bib}

\appendix
\clearpage
\section*{Appendix}
\renewcommand{\thesection}{\Alph{section}}
\renewcommand{\thetable}{A.\arabic{table}}
\setcounter{table}{0}

\begin{table*}[htbp]
    \centering
    \caption{Absolute total energy differences in $E_\mathrm{h}$ for DFT calculations using different codes, compared to \eminus{}.
    All calculations use the SVWN5 exchange-correlation functional, a cut-off energy of $30\,E_\mathrm{h}$, and a total energy convergence threshold of $10^{-8}\,E_\mathrm{h}$.
    For solids, the $\boldsymbol k$-point grid is given in parentheses.}
    \begin{tabular}{lccc}
        \toprule
        System & JDFTx & PWDFT.jl & QE \\
        \midrule
        \multicolumn{4}{c}{Molecules} \\
        \midrule
        H$_2$                      & $3.233\times10^{-6}$ & $1.518\times10^{-10}$           & $1.015\times10^{-8}$ \\
        N$_2$                      & $8.500\times10^{-8}$ & $1.628\times10^{-8\phantom{1}}$ & $6.518\times10^{-8}$ \\
        SiH$_4$                    & $6.014\times10^{-6}$ & $7.811\times10^{-10}$           & $7.481\times10^{-9}$ \\
        Benzene                    & $9.883\times10^{-6}$ & $1.673\times10^{-9\phantom{1}}$ & $9.633\times10^{-8}$ \\
        \midrule
        \multicolumn{4}{c}{Solids} \\
        \midrule
        Si (1$\times$1$\times$1)   & $2.095\times10^{-8}$ & $1.421\times10^{-8\phantom{1}}$ & $2.961\times10^{-8}$ \\
        Si (5$\times$5$\times$5)   & $2.652\times10^{-8}$ & $9.502\times10^{-9\phantom{1}}$ & $3.210\times10^{-8}$ \\
        GaAs (1$\times$1$\times$1) & $5.451\times10^{-8}$ & $5.848\times10^{-8\phantom{1}}$ & $7.768\times10^{-8}$ \\
        GaAs (5$\times$5$\times$5) & $4.444\times10^{-8}$ & $3.425\times10^{-8\phantom{1}}$ & $4.775\times10^{-8}$ \\
        \bottomrule
    \end{tabular}
    \label{tab:energies}
\end{table*}
\clearpage

\section{Benchmark timings}
\label{sec:timings}
To allow comparable benchmarks between conceptionally different electronic structure codes, various numerical parameters need to be chosen cautiously.
In addition to the mentioned computational parameters in Sec. \ref{subsec:benchmarks}, we ensured in all codes the usage of the same SVWN5 exchange-correlation functional, GTH pseudopotentials, identical $\boldsymbol k$-point sampling, and initial wavefunctions generated from random values.

These calculations, as seen in Fig. \ref{fig:energies}, have been timed, performed on an AMD Ryzen 5 5500u CPU with 16\,GB RAM using 6 OpenMP threads.
The total calculation time has been measured, along with the time per SCF cycle (calculated by dividing the total calculation time by the number of SCF cycles).
The results with the calculated speedup of the total calculation time over \eminus{} are listed in Tab. \ref{tab:timings}.
It should be noted that these timings only serve as a general overview.
They neglect that, e.g., JDFTx and QE offer parallelization over multiple nodes and GPU support.

One can see that for $\Gamma$-point calculations, the performance of \eminus{} is close to the other codes.
The calculation time is lower than for PWDFT.jl (the precompilation step from Julia has been excluded from all timings) while JDFTx and QE are about twice as fast (both in the total calculation time and time per SCF cycle).
While Python is generally considered slow, the most expensive step in these calculations is the evaluation of fast Fourier transformations (FFTs).
Using NumPy/SciPy/Torch, the FFT operations use optimized BLAS routines.
The implementation of the commonly employed FFTW library in the form of pyFFTW \cite{Gomersall2016} did not show any performance improvements.

Larger timing differences can be seen for finer $\boldsymbol k$-point grids, where the other codes give a speedup of factor 2 to 25 for the given calculations.
The calculation over multiple $\boldsymbol k$-points can easily be parallelized, where only the energy evaluation needs the information of all processes.
However, the parallelization of Python is rather cumbersome, often resulting in slower code compared to using the intrinsic parallelization of NumPy/SciPy while making the code less readable at the same time.
This is why no explicit $\boldsymbol k$-point parallelization has been implemented yet.
However, this might change in the future as recent developments provide more tools to mitigate this problem.
Noteworthy examples are the experimental removal of the global interpreter lock in Python 3.13 or modern applications like Ray \cite{Moritz2018}.

\begin{table*}[htbp]
    \centering
    \caption{Calculation time, number of SCF cycles, time per SCF cycle, and speedup over \eminus{} for DFT calculations using different codes for the calculations in Tab. \ref{tab:energies}.}
    \scalebox{0.94}{
    \begin{tabular}{lcccc}
        \toprule
        System & \eminus{} & JDFTx & PWDFT.jl & QE \\
        \midrule
        \multicolumn{5}{c}{Calculation time [s]} \\
        \midrule
        H$_2$                      & \phantom{11}7.47 & \phantom{11}5.59 & \phantom{1}12.37 & \phantom{11}3.48 \\
        N$_2$                      & \phantom{1}25.78 & \phantom{1}11.67 & \phantom{1}43.07 & \phantom{11}9.89 \\
        SiH$_4$                    & \phantom{1}17.60 & \phantom{11}8.72 & \phantom{1}25.14 & \phantom{1}12.55 \\
        Benzene                    & \phantom{1}65.59 & \phantom{1}27.37 & \phantom{1}96.36 & \phantom{1}25.67 \\
        Si (1$\times$1$\times$1)   & \phantom{11}3.06 & \phantom{11}1.61 & \phantom{11}2.36 & \phantom{11}1.08 \\
        Si (5$\times$5$\times$5)   & 243.23 & \phantom{1}47.28 & \phantom{1}57.06 & \phantom{1}11.00 \\
        GaAs (1$\times$1$\times$1) & \phantom{11}9.05 & \phantom{11}4.73 & \phantom{11}3.77 & \phantom{11}5.02 \\
        GaAs (5$\times$5$\times$5) & 450.67 & 183.64 & \phantom{1}97.70 & \phantom{1}17.58 \\
        \midrule
        \multicolumn{5}{c}{Number of cycles} \\
        \midrule
        H$_2$                      & \phantom{1}12\phantom{.11} & \phantom{1}14\phantom{.11} & \phantom{1}12\phantom{.11} & \phantom{11}5\phantom{.11} \\
        N$_2$                      & \phantom{1}25\phantom{.11} & \phantom{1}25\phantom{.11} & \phantom{1}28\phantom{.11} & \phantom{11}9\phantom{.11} \\
        SiH$_4$                    & \phantom{1}18\phantom{.11} & \phantom{1}17\phantom{.11} & \phantom{1}17\phantom{.11} & \phantom{1}15\phantom{.11} \\
        Benzene                    & \phantom{1}28\phantom{.11} & \phantom{1}27\phantom{.11} & \phantom{1}30\phantom{.11} & \phantom{1}10\phantom{.11} \\
        Si (1$\times$1$\times$1)   & \phantom{1}40\phantom{.11} & \phantom{1}30\phantom{.11} & \phantom{1}33\phantom{.11} & \phantom{1}11\phantom{.11} \\
        Si (5$\times$5$\times$5)   & \phantom{1}36\phantom{.11} & \phantom{1}24\phantom{.11} & \phantom{1}24\phantom{.11} & \phantom{11}6\phantom{.11} \\
        GaAs (1$\times$1$\times$1) & 119\phantom{.11} & \phantom{1}75\phantom{.11} & \phantom{1}53\phantom{.11} & 117\phantom{.11} \\
        GaAs (5$\times$5$\times$5) & \phantom{1}52\phantom{.11} & \phantom{1}38\phantom{.11} & \phantom{1}34\phantom{.11} & \phantom{11}6\phantom{.11} \\
        \midrule
        \multicolumn{5}{c}{Time per SCF cycle [s]} \\
        \midrule
        H$_2$                      & \phantom{11}0.62 & \phantom{11}0.40 & \phantom{11}1.03 & \phantom{11}0.70 \\
        N$_2$                      & \phantom{11}1.03 & \phantom{11}0.47 & \phantom{11}1.54 & \phantom{11}1.10 \\
        SiH$_4$                    & \phantom{11}0.98 & \phantom{11}0.51 & \phantom{11}1.48 & \phantom{11}0.84 \\
        Benzene                    & \phantom{11}2.34 & \phantom{11}1.01 & \phantom{11}3.21 & \phantom{11}2.57 \\
        Si (1$\times$1$\times$1)   & \phantom{11}0.08 & \phantom{11}0.05 & \phantom{11}0.07 & \phantom{11}0.10 \\
        Si (5$\times$5$\times$5)   & \phantom{11}6.76 & \phantom{11}1.97 & \phantom{11}2.38 & \phantom{11}1.83 \\
        GaAs (1$\times$1$\times$1) & \phantom{11}0.08 & \phantom{11}0.06 & \phantom{11}0.07 & \phantom{11}0.04 \\
        GaAs (5$\times$5$\times$5) & \phantom{11}8.67 & \phantom{11}4.83 & \phantom{11}2.87 & \phantom{11}2.93 \\
        \midrule
        \multicolumn{5}{c}{Speedup} \\
        \midrule
        H$_2$                      & \phantom{11}1.0\phantom{1} & \phantom{11}1.3\phantom{1} & \phantom{11}0.6\phantom{1} & \phantom{11}2.1\phantom{1} \\
        N$_2$                      & \phantom{11}1.0\phantom{1} & \phantom{11}2.2\phantom{1} & \phantom{11}0.6\phantom{1} & \phantom{11}2.6\phantom{1} \\
        SiH$_4$                    & \phantom{11}1.0\phantom{1} & \phantom{11}2.0\phantom{1} & \phantom{11}0.7\phantom{1} & \phantom{11}1.4\phantom{1} \\
        Benzene                    & \phantom{11}1.0\phantom{1} & \phantom{11}2.4\phantom{1} & \phantom{11}0.7\phantom{1} & \phantom{11}2.6\phantom{1} \\
        Si (1$\times$1$\times$1)   & \phantom{11}1.0\phantom{1} & \phantom{11}1.9\phantom{1} & \phantom{11}1.3\phantom{1} & \phantom{11}2.8\phantom{1} \\
        Si (5$\times$5$\times$5)   & \phantom{11}1.0\phantom{1} & \phantom{11}5.1\phantom{1} & \phantom{11}4.3\phantom{1} & \phantom{1}22.1\phantom{1} \\
        GaAs (1$\times$1$\times$1) & \phantom{11}1.0\phantom{1} & \phantom{11}1.9\phantom{1} & \phantom{11}2.4\phantom{1} & \phantom{11}1.8\phantom{1} \\
        GaAs (5$\times$5$\times$5) & \phantom{11}1.0\phantom{1} & \phantom{11}2.5\phantom{1} & \phantom{11}4.6\phantom{1} & \phantom{1}25.6\phantom{1} \\
        \bottomrule
    \end{tabular}
    }
    \label{tab:timings}
\end{table*}

\end{document}